# A new macroscopically degenerate ground state in the spin ice compound $Dy_2Ti_2O_7$ under a magnetic field

K. Matsuhira[1], Z. Hiroi[2], T. Tayama[2], S. Takagi[1] and T. Sakakibara[2]

[1] Department of Electronics, Faculty of Engineering, Kyushu Institute of Technology,

Kitakyushu 804-8550, Japan

[2] Institute for Solid State Physics, University of Tokyo, Kashiwa 277-8581, Japan

The low temperature magnetic properties of pyrochlore compound $Dy_2Ti_2O_7$ in magnetic fields applied along the [111] direction are reported. Below 1 K, a clear plateau has been observed in the magnetization process in the field range 2~9 kOe, followed by a sharp moment jump at around 10 kOe that corresponds to a breaking of the spin ice state. We found that the plateau state is disordered with the residual entropy of nearly half the value of the zero-field state, whose macroscopic degeneracy comes from a frustration of the spins on the kagomé layers perpendicular to the magnetic field.

E-mail: matuhira@elcs.kyutech.ac.jp (K. Matsuhira)

Geometrical frustration can lead to novel phenomena such as a macroscopic degeneracy in the ground state with no long-range ordering. Recently, pyrochlore oxides have been attracting much interest because the structure, including corner-shared tetrahedra whose vertices are occupied by spins (figure 1a), may show a strong frustration [1-8]. In the pyrochlore lattice, even a ferromagnetic coupling between spins can lead to frustration [4,8,9] when there is a strong single-site anisotropy along the local <111> axes. $Dy_2Ti_2O_7$ is the typical example of this case, where the ferromagnetic interaction stabilizes the local spin arrangement of two spins pointing outward and two spins inwards (so-called `two-in two-out' state) in a basic tetrahedron (figure 1a). For every tetrahedron, there are six possible combinations of spins under the two-in two-out rule reflecting the global cubic symmetry. Because the ground state is highly degenerate, a static disordered state (so-called ``spin ice" state) is formed below 1 K in spite of structurally ordered system. In fact, $Dy_2Ti_2O_7$ is found to show the residual ground state entropy of 1.68~J/mol K, which is numerically in agreement with the Pauling's entropy for water ice [10,11].

The spin ice state is closely related to the high symmetry of the pyrochlore structure. Novel features might therefore be expected in the Ising pyrochlore ferromagnet when the degeneracy is partially or totally reduced. Applying magnetic field is an interesting and the easiest method to lower the symmetry, and new types of phase transitions are predicted, depending on the field direction [12] In this letter, we report the magnetic properties of pyrochlore compound $Dy_2Ti_2O_7$ in magnetic fields applied along the [111] direction. The pyrochlore lattice can be viewed as an alternating stacking of kagomé layers and sparse triangular layers, both of which are perpendicular to the [111] direction (figure 1b). By applying magnetic field along the [111] direction, we found a new macroscopically degenerate state in which, although the magnetization is induced, a macroscopic degeneracy remains in the spin configuration on the kagomé layers.

Single crystals of $Dy_2Ti_2O_7$ were prepared by the floating-zone method using an infrared furnace equipped with four halogen lamps and elliptical mirrors. The crystals were grown under $O_2$ gas flow to avoid oxygen deficiency. The typical growth rate was 4 mm/h. The obtained single crystals were translucent yellow. DC magnetization measurements were done by a capacitive Faraday magnetometer installed in a $^3$He cryostat, with a field gradient of 300 Oe/cm [13]. Specific heat measurements were carried out by a relaxation method (PPMS, Quantum Design) down to 0.4 K. The typical sizes of the samples used for magnetization and specific heat measurements were $0.5 \times 2 \times 2$ mm$^3$ and $0.1 \times 1 \times 1$ mm$^3$, respectively. In order to minimize the demagnetizing field effect, the [111] direction was oriented along the sample plane.

DC magnetization curves of $Dy_2Ti_2O_7$ with magnetic field applied along the [111] direction are shown in figure 2. At $T$=1.65 K, the magnetization is a gradual function of field with a weak feature at around 10 kOe, and saturates at higher fields to the value ~ 5 $\mu_B$/Dy. This moment value corresponds to the fully saturated one-in three-out (or three-in one-out) state of the Ising pyrochlore lattice with the local Ising axis pointing along the <111> directions. The result at 1.65 K is in good agreement with the previous measurements done at 1.8 K by Fukazawa *et al*. [14].

On cooling below 1 K, the feature at ~10 kOe becomes sharper, and eventually turns into a metamagnetic step at $T$=0.48 K with a preceding magnetization plateau below ~9 kOe. The magnetic moment of the plateau is very close to the value 3.33 $\mu_B$/Dy, expected for the saturated moment along the [111] direction without destroying the ice rule (two-in two-out state). In fact, the emergence of the plateau is closely associated with a formation of the spin ice state which is evidenced by an appearance of the magnetization hysteresis below 5 kOe [7]. Clearly, the metamagnetic step near 10 kOe corresponds to a breaking of the spin ice state by a magnetic field strong enough to overcome the magnetic interactions. This phenomenon has been predicted by Monte Carlo simulations of the spin ice models [12,14,15], but experimentally only a broad feature had been observed in the previous measurements done at 1.8 K [14]. Our data are the first results in the low temperature regime where the ice rule configuration is well developed. Magnetization measurements at still lower temperatures are in progress and the results will be published soon [16].

In the plateau state, each spin on the triangular layers is considered to be aligned since its Ising axis is parallel to the field, rendering the Zeeman energy the largest. The remaining spins on the kagomé layers, however, can only be partially ordered because the ice rule is maintained. This suggests a new macroscopically degenerate state in this intermediate field range, as will be confirmed by the specific heat measurements. We could fit the raw data taken at zero field in the temperature range between 12 K and 19 K to a conventional form for lattice contribution $C=\alpha T^3$ with $\alpha=4.85 \times 10^{-4}$ J/K$^4$ mole-Dy, which suggests that magnetic contribution is almost absent there.

Assuming that this lattice contribution is independent of magnetic fields, magnetic specific heat at low temperature below 15 K was obtained by subtracting it from the raw data. On the other hand, magnetic specific heat at higher temperature was estimated by subtracting the zero-field data as a lattice part. However, this estimation may suffer from increasing experimental error with increasing temperature, because the lattice part becomes dominant. In figure 3, we show the temperature dependence of the magnetic specific heat $C(T)$ of $Dy_2Ti_2O_7$ under various fields applied along the [111] direction. At zero filed, $C(T)$ shows a broad maximum at 1.2 K which is presumably related to a freezing process to the highly degenerate spin ice state. Applying a magnetic field of 5 kOe, $C(T)$ shows not only a broad maximum at 1.2 K but also a shoulder at 3 K. These features of $C(T)$ split into two broad peaks at the field of 7.5 kOe. The higher temperature peak at ~ 4 K shifts to still higher temperature side with further increasing field. This peak probably arises from the Schottky anomaly of Zeeman split spins on the triangular layers, which are parallel to the field direction and more stabilized with increasing fields.

On the other hand, the origin of the lower temperature peak in $C(T)$, which may come from the spins on the kagomé layer, seems to be more complicated. These spins have a component of the magnetic moment along the [111] field direction 1/3 times smaller than that on the triangular layers and are therefore subjected to a smaller Zeeman energy. Interestingly, this broad peak shifts to lower temperature side with increasing field up to 10 kOe. This fact implies that the lower temperature peak arises from certain short-ranged magnetic correlations in the kagomé layer, which must have a close relation to the plateau state observed in the magnetization. The peak becomes very sharp at around 10 kOe, and shifts again to a higher temperature side with further increasing fields. In the strong magnetic field regime above 10 kOe, we observe two broad peaks in $C(T)$ moving towards the high temperature side, both of which are ascribed to the Schottky type anomalies from two types of spins on the kagomé and the triangular layers with different Zeeman splittings.

In order to explore the nature of the plateau state further, we estimated the field variation of the magnetic entropy $S(T)$ by integrating $C/T$ in magnetic fields. We make an assumption that $S(T)$ approaches the value $R\ln 2$ expected for a Kramers doublet of the Dy ions, where $R$ is the gas constant, at temperatures much higher than the ground state manifolds but still lower than the range where the crystal-field split higher levels are populated. As can be seen from figure 3, the zero field $C(T)$ value above 10 K falls off to nearly zero, indicating the populations of the excited levels are negligible at this temperature range. We can thus estimate the absolute entropy value at various fields. Figure 4 shows the field dependence of the magnetic entropy $S(H)$ estimated at 0.4 K. Since the zero-field $C(T)$ already dies out at this temperature, the $S(H=0)$ value of ~1.6 J/mole-Dy at 0.4 K would be a good estimate of the ground state entropy. This value is very close to the residual magnetic entropy of the spin ice state ($R$ (1/2)ln(3/2)=1.68 J/K mole), and confirms that the system is in the macroscopically degenerate state at zero field. This residual entropy is partially released by applying magnetic field of 2 kOe, because the sixfold degenerate states of the basic tetrahedron become energetically inequivalent. However, the magnetic entropy at 0.4 K still retains a finite value of $0.8 \pm 0.1$ J/K mole-Dy in the field range of 2.5 ~ 7.5 kOe where the magnetization plateau is observed. Because the $C(T)$ value at 5 kOe drops by almost two orders of magnitude on cooling below 1 K to 0.4 K, further entropy release is unlikely to occur below 0.4 K. Therefore, we conclude that the plateau state has a residual ground state entropy of $0.8 \pm 0.1$ J/K mole-Dy.

The specific heat data indicate that the $C(T)$ value at 10 kOe becomes very large at 0.4 K, implying a rather steep entropy release should occur at this field region, as can be expected for the breaking of the ice rule by magnetic field. This feature can be seen in figure 4 as a weak step in the $S(H)$ value at around 10 kOe. At fields well above 15 kOe, the spin configuration in the tetrahedron is the unique one-in three-out state. The system should have no residual ground-state entropy at all. In our estimate, however, $S(H)$ above 15 kOe still remains finite to the value of ~ 0.4 J/K mole-Dy. The reason for this offset might be that the Schottky peak of $C(T)$ above 16 kOe has a long tail to higher $T$ side much above 20 K; an experimental error would then be enhanced by the integration of $C/T$ towards higher temperature range and we are probably underestimating the entropy change above 20 K. More precise measurements would be needed to clarify this point.

Let us discuss the spin configuration in the magnetization plateau state. The magnetic field along the [111] direction reduces the sixfold degenerate two-in two-out configurations in the basic tetrahedron to threefold. In this state, every spin on the triangular layer is oriented along the [111] direction whereas the spins on the kagomé layer are still frustrated because the two-in two-out ice rule is maintained in each tetrahedron. The kagomé layer consists of corner-shared triangles (see figure 1b), and under the ice rule, there are two types of triangles having the spin configurations of either one-in two-out or two-in one-out. For each type, there are three ways to align the spins as shown in figure 5. Here a, b and c indicate the spin configurations of one-in two-out states in the triangles pointing downward, whereas d, e and f are those of two-in one-out states in the upward triangles. Clearly, the plateau state is frustrated and may have a residual entropy which is recently calculated to be about 40 % of the zero-field spin ice state [17], in reasonably good agreement with the results in figure 4. Our experimental data of the magnetization and the specific heat measurements thus confirm this new macroscopically degenerate state: `kagomé ice' state in $Dy_2Ti_2O_7$ under magnetic fields.

We would like to thank K. Nemoto for useful discussions. This work was carried out under the Visiting Researcher's Program of the Institute for Solid State Physics, the University of Tokyo. Crystal growth and specific heat measurement were performed using facilities of the Materials Design and Characterization Laboratory, Institute for Solid State Physics, University of Tokyo. This work was also partly supported by a Grant-in-Aid for Scientific Research of the Ministry of Education, Culture, Sports, Science and Technology of Japan.

**Figure Captions**

**Figure 1**

a, Pyrochlore lattice of the corner-sharing tetrahedra whose vertices are occupied by Ising spins of Dy ions. The long allow shows the [111] direction along which the magnetic field is applied. One of the six two-in, two-out configurations is shown by short arrows. Each spin lies along the axis joining the vertex and the center of the tetrahedron, due to a strong single-site anisotropy. b, Pyrochlore lattice viewed along the [111] direction. It consists of an alternating stacking of kagomé and triangular layers.

**Figure 2**

Magnetization process of $Dy_2Ti_2O_7$ along the [111] direction.

**Figure 3**

Temperature dependence of the magnetic specific heat $C$ of $Dy_2Ti_2O_7$ measured at various magnetic fields applied along the [111] direction.

**Figure 4**

Magnetic field dependence of magnetic entropy $S(H)$ at 0.4 K. Dashed-dotted line shows a residual ground state entropy of spin ice ($R(1/2)\ln(3/2)$). $S(H)$ shows a plateau (broken line) in the magnetic field range of 3.5 ~ 7 kOe.

**Figure 5**

Macroscopic degeneracy in the plateau state (kagomé ice). **a**, **b** and **c** show the three patterns for the one-in two-out configurations for triangles with another apical spins above, whereas **d**, **e** and **f** are those for two-in one-out arrangements for the other triangles with another apical spins below in the kagomé layer.

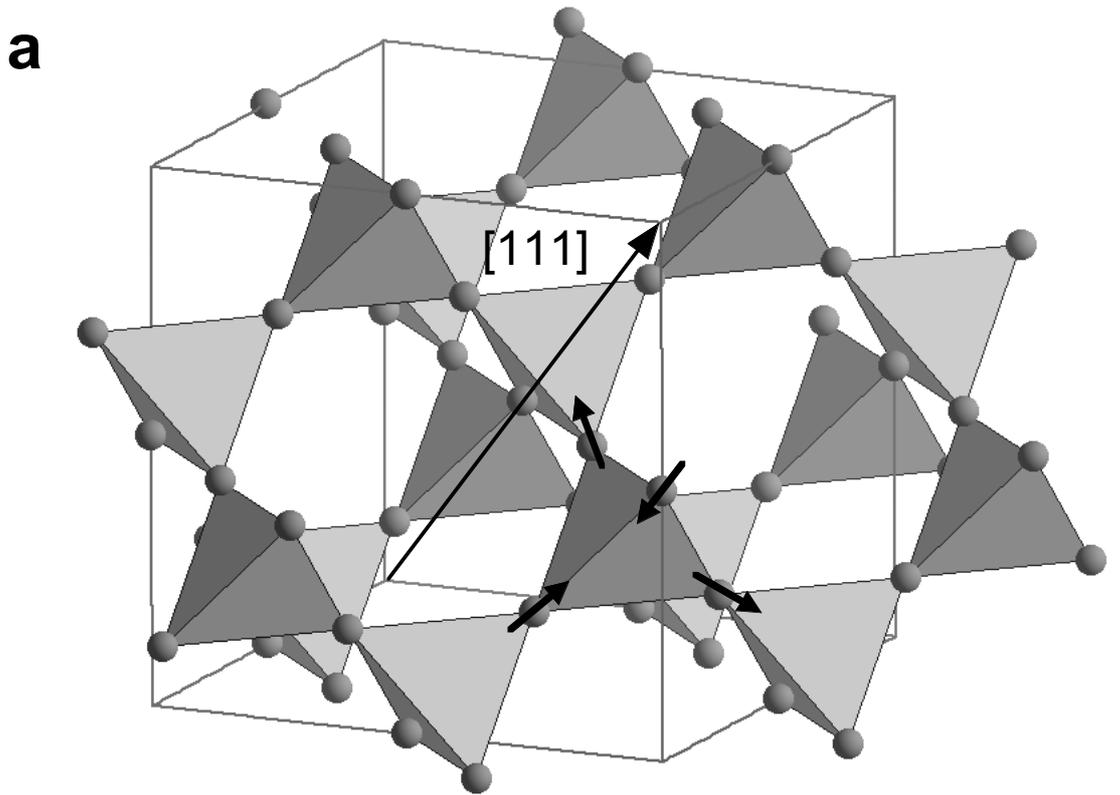

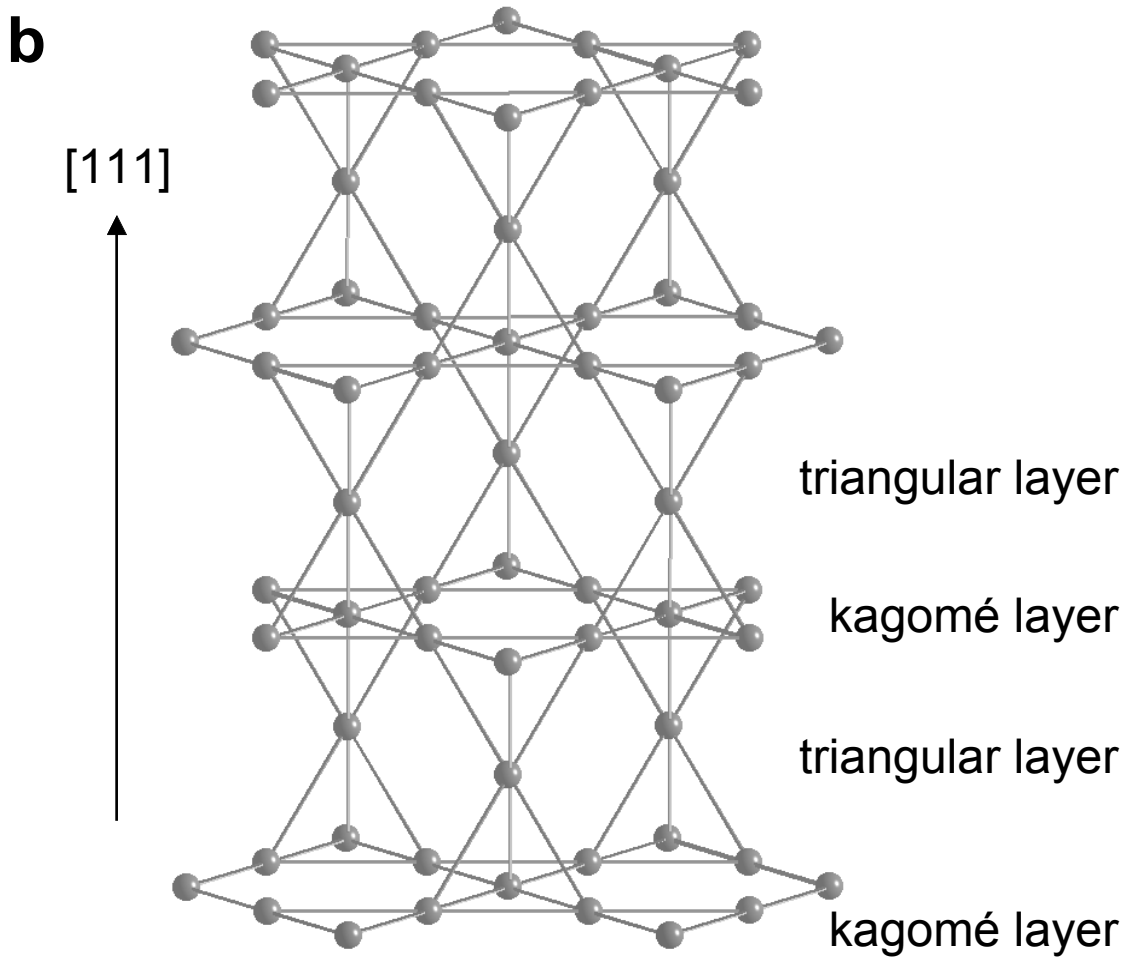

**Figure 1**

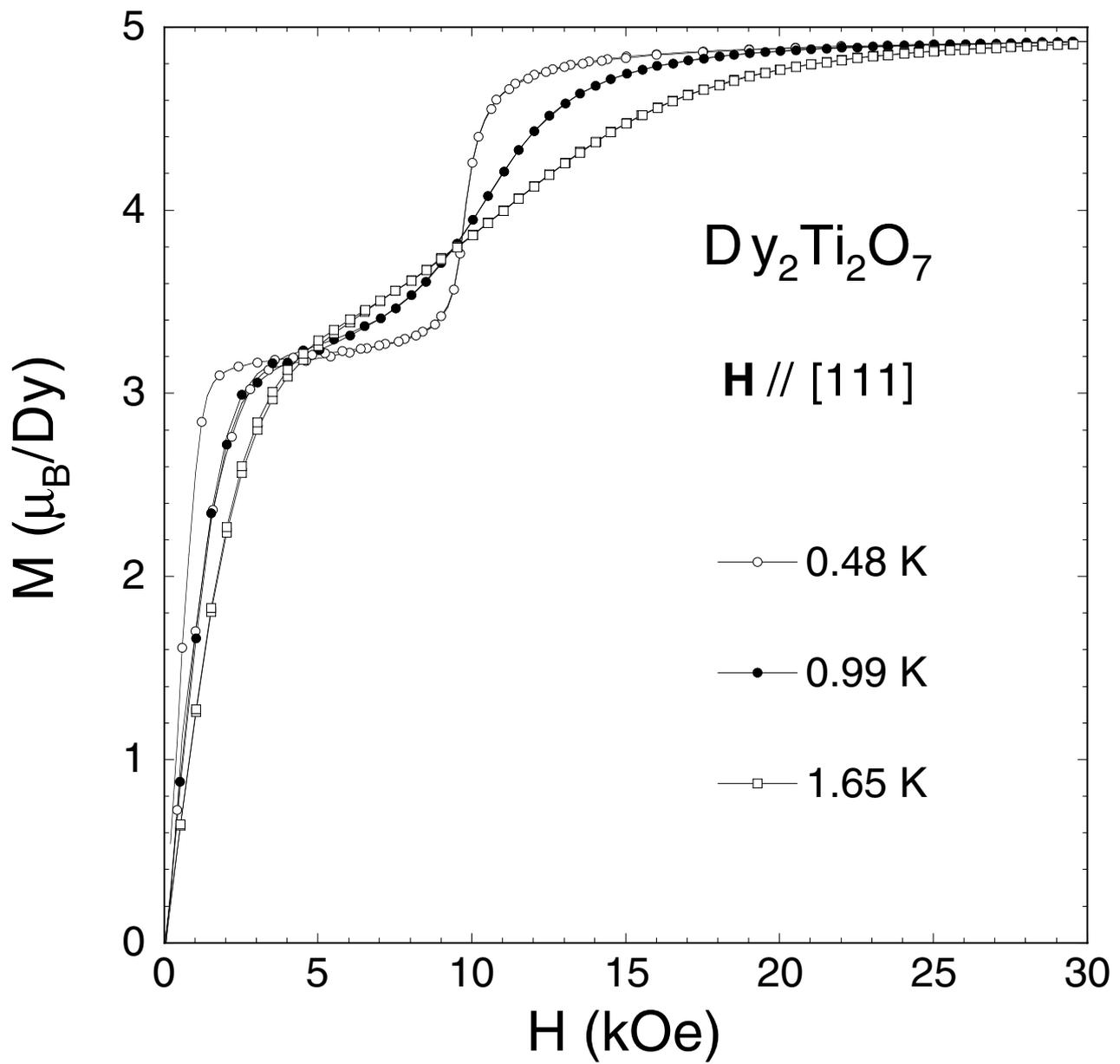

**Figure 2**

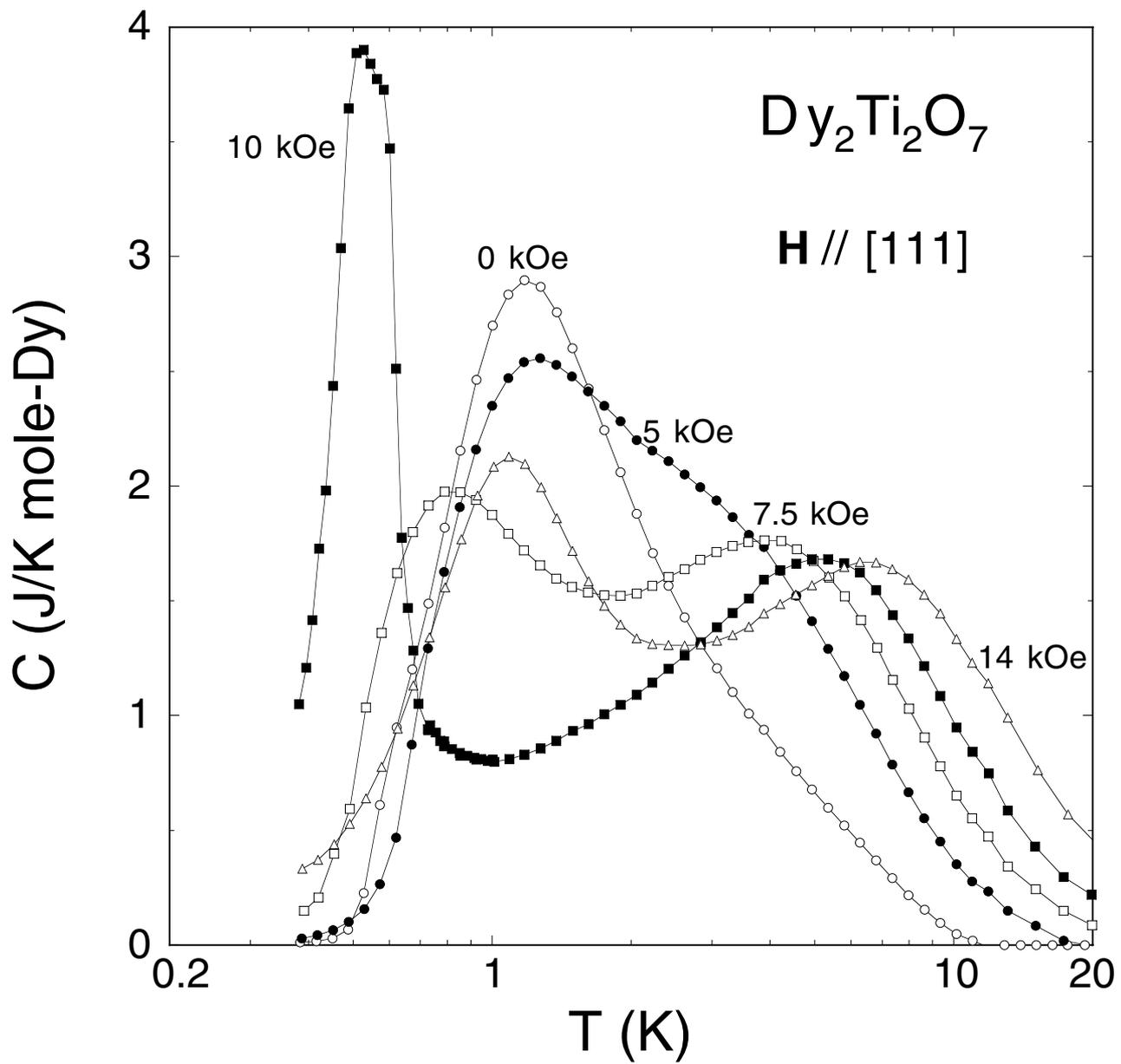

Figure 3

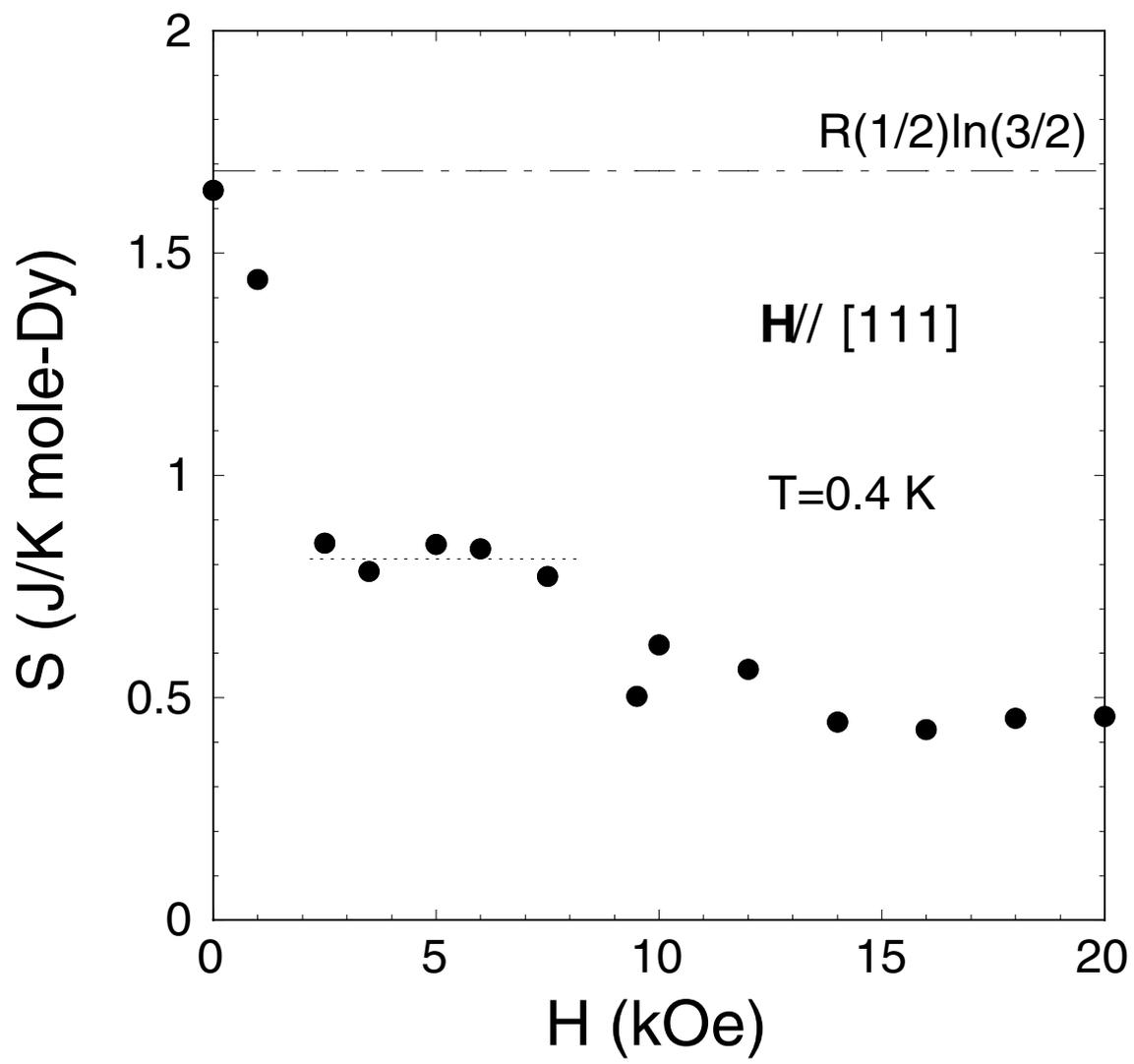

**Figure 4**

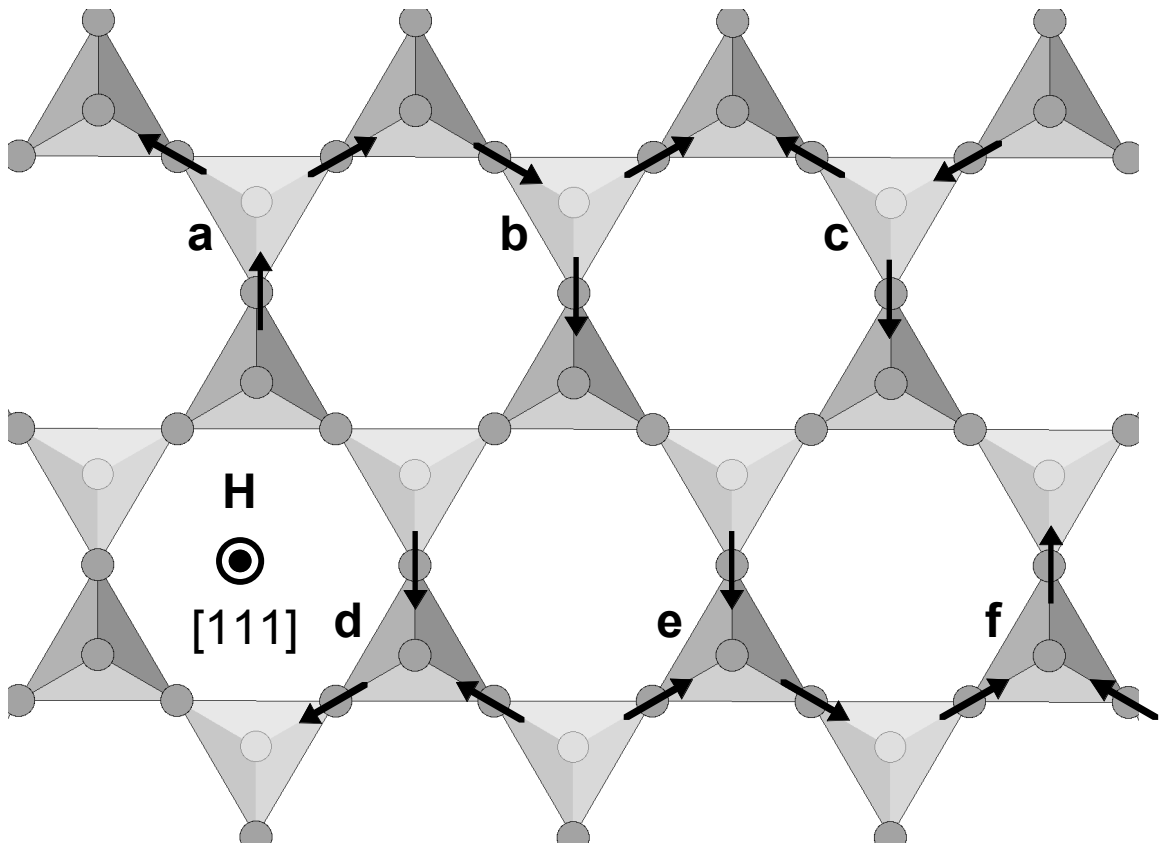

**Figure 5**